\documentstyle[prb,aps,epsf]{revtex}
\begin{document}
\draft
\title{Heat Capacity in Bits}
\author{P. Fraundorf}
\address{Physics \& Astronomy, U. Missouri-StL (63121) \\
Physics, Washington U. (63110), \\
St. Louis, MO, USA}
\date{\today }
\maketitle

\begin{abstract}
Information theory this century has clarified the 19th century work of
Gibbs, and has shown that natural units for temperature kT, defined via $%
1/T\equiv dS/dE$, are energy per nat of information uncertainty. This means
that (for {\em any} system) the total thermal energy $E$ over $kT$ is the
log-log derivative of multiplicity with respect to energy, and (for {\em all} b)
the number of base-$b$ units of information lost about the state of the
system per $b$-fold increase in the amount of thermal energy therein. For
``un-inverted'' ($T>0$) systems, $E/kT$ is also a temperature-averaged
heat capacity, equaling ``degrees-freedom over two'' for the quadratic case.
In similar units the work-free differential heat capacity $C_v/k$ is a
``local version'' of this log-log derivative, equal to bits of uncertainty
gained per 2-fold increase in temperature. This makes $C_v/k$ (unlike $E/kT$%
) independent of the energy zero, explaining in statistical terms its 
usefulness for detecting both phase changes and quadratic modes. From 
UMStL-CME-94a09pf.
\end{abstract}


%\onecolumn

%\twocolumn

\section{Introduction}

Those conceptual approaches to a subject that offer both wider
applicability, and reduced algorithmic complexity, might be called ``deep
simplifications''. Minkowski's approach to special relativity via the metric
equation is a classic example. The first impressions of ``superfluous
erudition''\cite{Einstein} have now been eclipsed, by uses for Minkowski's
perspective in developing general relativity\cite{GR} as well as first
introductions to space-time\cite{Taylor,Bell,Fraundorf}.

Similarly, the communication-theory insights of Shannon\cite{Shannon} led
Jaynes\cite{Jaynes} in the middle of this century to clarify the distinction
in thermal physics between ``describing the dice'' (i.e. physical
description), and ``taking the best guess'' (the ``gambling theory'' part).
Thus for example, the 1st through 3rd laws of thermodynamics help ``describe
the dice'', while the zeroth law, as well as the Boltzmann (canonical) and
Gibbs (grand canonical) factors, are ``dice-independent'' tools of
statistical inference\cite{Plischke}. Hence we can (with increasing help
from undergraduate texts\cite
{Girifalco,Kittel,Stowe,Betts,Garrod,Schroeder,Serway}) deepen understanding
(including the physical intuition of computer science and biology students
on matters of ``code'') by disclosing that entropy $S$ is a measure of
statistical uncertainty, expressible using information units as well as in
J/K. From this, it follows that temperature (i.e. the reciprocal of
uncertainty slope $1/T\equiv dS/dE$) can be measured in units of energy per
unit information (e.g. room temperature $\approx $ $\frac 1{40}$ eV per
nat). Thus heat capacities (i.e. $dE/dT$), normally understood in units of
energy per degree Kelvin, may find themselves measured in units of
information alone! As we will show, this is part of a larger trend in
statistical physics to shift the focus from temperature (or its reciprocal)
to the physical dependence of multiplicity (or entropy) on variables
conserved in the interaction between complex systems.

But, what is the {\em physical} meaning of a heat capacity without reference
to historical temperature units? Bits of what? An answer to this question
(something any student might ask) does not appear to be common knowledge
among physics teachers, so we outline an answer here. We show further why
total thermal energy over kT {\em for any system} serves at once: (i) as the
instantaneous exponent of energy in the expression for accessible state
multiplicity, (ii) as a measure of the number of bits of (micro-canonical)
uncertainty added per two-fold increase in thermal energy, and (iii) {\em %
for quadratic systems} as the number of degrees freedom over two.

\section{Describing the Dice}

The dice of thermal physics are usually physical systems capable of
accomodating thermal energy (as well as other quantities that may be
conserved, like volume and particles) in a {\em multiplicity} of ways. This
multiplicity is itself the key to understanding, particularly when systems
are seen from the (micro-canonical) vantage point of the conserved
quantities {\em per se}. For example, many gases, liquids, and solids behave
over some of their temperature range as though multiplicity ($\Omega $) is
proportional to $E^{\frac{\nu N}2}$, where $E$ is thermal energy, $N$ is the
number of molecules, and $\nu $ is the number of ways to store thermal
energy per molecule. Such systems are called ``quadratic'', since the
proportionality results from a ``sum of squares'' connection between $E$ and
the state coordinates involved in storing energy.

\section{Taking the Best Guess}

Application of gambling theory to physical systems begins with the question:
Where might we expect to find a conserved quantity $X$ which has been
randomly (as far as we know) shared between systems for so long that prior
information about the whereabouts of $X$ is irrelevant? In the jargon of the
field, this is the same as asking: Where is $X$ likely to be {\em after
equilibration}?

The science of decision-making in the presence of uncertainty (i.e.
statistical inference or ``gambling theory'') suggests that the best bet is
the distribution of $X$ that can happen in the most ways, provided there is
no reason to prefer one way over another. In the jargon of probability
theory, this is the recommendation: Assume equal {\em a priori}
probabilities, when evidence to the contrary is not available.

For example, if systems A and B have a total energy $E$ to share between
them, then the best bet after equilibration will be that value of $E_A=E-E_B$
that has the largest total multiplicity $\Omega =\Omega _A\Omega _B$.
Setting to zero the derivative of $\Omega $ with respect to $E_A$, one finds
that this maximum requires that $\frac 1{\Omega _A}\frac{d\Omega _A}{dE_A}=%
\frac 1{\Omega _B}\frac{d\Omega _B}{dE_B}$, and hence that $\frac{d(\ln
\Omega )}{dE}$ be the same for both systems! These derivatives (and most
others in this paper) are taken under ``microcanonical constraints'', i.e.
they are partial derivatives with other extensive quantities (like volume or
number of particles) held constant.

Information theorists, of course, define the logarithm of multiplicity as
uncertainty via the equation $S=k\ln \Omega $, and measure it in \{nats,
bits, bytes, or J/K\} if $k$ is \{$1$,$\frac 1{\ln 2}$,$\frac 1{\ln 256}$,$%
1.38\times 10^{-23}$\}, respectively. Thus the best bet for any two systems
sharing a conserved quantity $X$, in the absence of information to the
contrary, is that $X$ will rearrange itself between the two systems until
each system's {\em uncertainty slope} ($\frac{dS}{dX}$) has reached a common
value. This is a quantitative version of the zeroth law of thermodynamics,
based purely in the science of statistical inference, which applies to {\em %
any systems} sharing conserved quantities.

Of course, when energy $E$ is the quantity shared randomly between systems,
the uncertainty slope $\frac{dS}{dE}$ is the reciprocal temperature or {\em %
coldness} $\frac 1T$. Hence temperature is a property which signals the
propensity of a system for sharing of energy thermally. For example,
calculating the uncertainty slope for quadratic systems from the
multiplicity given above yields the widely useful equipartition relation: $%
\frac EN=\frac \nu 2kT$.

When $V$ is the quantity shared randomly between systems, the uncertainty
slope $\frac{dS}{dV}$ is the {\em free-expansion coefficient} equal to $%
\frac{dS}{dE}\frac{dE}{dV}=\frac 1T\frac{Fdx}{Adx}=\frac PT$ at equilibrium%
\cite{Garrod}. For an ideal gas, $\Omega \propto V^N$. Solving this for $%
\frac{dS}{dV}$ yields the ideal gas equation of state $PV=NkT$. When $N$ is
the quantity shared randomly, the uncertainty slope $\frac{dS}{dN}$ is the 
{\em chemical affinity}, equal to $\frac{-\mu }T$ at equilibrium. From this,
for example, reaction equilibrium constants may be calculated.

This quantitative version of the zeroth law applies to all thermal systems
which equilibrate, including spin systems (like magnets) capable of
population inversions and hence negative absolute temperatures. Moreover, as
a theorem of statistical inference not involving energy at all, it applies
also to thermally unequilibrated systems sharing other conserved quantities
(even money, for example), provided the only prior information we have is
how the multiplicity of ways that quantity can be distributed depends on the
amount of that conserved quantity to begin with! If we have other kinds of
information, such as knowledge of a system's temperature but not its total
energy, then the broader class of maximum entropy strategies in statistical
inference (e.g. the canonical and grand ensembles) predict the distribution
of outcomes we can expect there as well.

\section{Thermal Energy over kT}

A closer look shows that the statistical definition of temperature above can
be rewritten as:

\begin{equation}
\frac 1{kT}\equiv \frac{\partial (\ln \Omega )}{\partial E}=\left\{ \frac{%
\partial (\ln \Omega )}{\partial \left( \ln E\right) }\right\} \frac 1E\text{%
.}  \label{temperature}
\end{equation}
The quantity in curly brackets is the log-log derivative of multiplicity
with respect to thermal energy. We can also think of this as the
``instantaneous exponent'' of energy in the expression of multiplicity as
energy to some power, or as the slope of the multiplicity versus energy
curve on a log-log plot.

Rearranging the equation yields something that looks very much like the
familiar equipartition theorem, except that the relation applies to all
thermal systems under conditions of maximum ignorance (i.e. at equilibrium):

\begin{equation}
\frac E{kT}=E\frac{\partial (S/k)}{\partial E}=E\frac{\partial \ln \Omega }{%
\partial E}\equiv \xi \stackrel{E>0}{=}\left\{ \frac{\partial \ln \Omega }{%
\partial \ln E}\right\} =\left\{ \frac{\partial (\log _b\Omega )}{\partial
\left( \log _bE\right) }\right\} \forall b\in \{R+\}.  \label{degreesfreedom}
\end{equation}
For quadratic systems, it is easy to see that our log-log derivative is
nothing more than half the number of degrees of freedom ($\frac{\nu N}2$).
For any system, however, $\xi $ measures the instantaneous energy exponent,
as well as the number of {\em nats of information lost about the state of
the system per e-fold increase in thermal energy} of the system.The last
term in the equality string simply notes that the value is independent of
the base $b$ of the logarithms used, provided the same base is used in
numerator and denominator. Thus we can also think of $\xi $ as the number of
bits of information lost per $2$-fold increase in thermal energy, or more
generally the number of base-$b$ units of information lost per $b$-fold
increase in thermal energy. In our search for the meaning of heat capacity
in natural units, this is our first big clue.

Before we move on, we should also point out something that applies if our
energy origin has been chosen so that $E\rightarrow 0$ as $T\rightarrow 0$,
something we {\em might} expect for a measure of thermal energy. In terms of
the no-work (e.g. constant volume) heat-capacity $C_v\equiv \frac{\partial E%
}{\partial T}$, we can write:

\begin{equation}
\xi =\frac E{kT}=\frac 1{kT}\int_0^TC_vdT\stackrel{T>0}{=}\frac 1{k\Delta T}%
\int_0^TC_vdT\equiv \left\langle \frac{C_v}k\right\rangle \text{,}
\label{integralcapacity}
\end{equation}
with the middle equality applying only for systems NOT in a population
inversion, so that absolute temperature $T>0$ and $\Delta T=T-0=T$. Thus
when absolute temperature is positive, $\xi $ is a {\em heat capacity average%
} over temperatures ranging between $T$ and absolute zero.

We've shown here that the log-log derivative of multiplicity, with respect
to energy, has a simple information theoretic interpretation, and is
elegantly given in natural units by $\frac E{kT}$ as well. From the
perspective of an experimentalist, however, it has one glaring disadvantage: 
{\em It's numeric value depends on our choice for the zero of thermal energy.%
}

To illustrate the problem, consider the cooling of water until it becomes
ice. As water cools initially, the temperature drop per unit energy removed
is roughly constant. One might easily say: ``This looks like a quadratic
system with about 18 degrees of freedom per molecule'', so $\xi $ must be
about 9 bits per 2-fold increase in thermal energy. Then, at the freezing
point, the temperature stops dropping as energy continues to be removed,
suggesting a quadratic system with nearly infinite degrees of freedom! Once
all is frozen, of course, temperature continues its drop, this time
suggesting a quadratic system with about 8 degrees of freedom per molecule,
or $\xi $ closer to 4 bits per 2-fold increase in thermal energy! Since $kT$
may change little during this experiment, how can $E/kT$ be jumping around
so much?

The answer of course is that these inference follow not by measuring total 
energy $E$, but only changes in energy. Moreover, in the process our
preferred zero of thermal energy has been shifting about. We can see the
effects of this more explicitly if we plot energy versus temperature for water, 
as shown in the lower left panel of Figure 1. 
The question then is, can we modify our estimate for the log-log derivative 
of thermal energy so as to reflect only data on temperature changes over a 
limited energy range? Such a quantity might allow us to probe the ways that 
thermal energy is being accomodated specifically, one energy range at a time.

\section{Instantaneous Heat Capacities}

The no-work (e.g. constant volume) instantaneous heat capacity, in natural
units, can be written in terms above as:

\begin{equation}
\frac{C_v}k\equiv \frac{\partial E}{k\partial T}=\frac \partial {\partial T}%
\left[ \xi T\right] =\left[ 1+T\frac \partial {\partial T}\right] \xi =T%
\frac{\partial \ln \Omega }{\partial T}=T\frac{\partial (S/k)}{\partial T}%
\text{.}  \label{heatcapacity}
\end{equation}
Here $T$ is absolute temperature in any units you like, and the partials are
taken with work parameters (like volume) held constant.

This quantity has an interesting property. If we define thermal energy $E$
as a difference between total energy $U$ and a specified ``zero thermal
energy'' origin $U_o$, i.e. as $E\equiv U-U_o$, then it is easy to see that $%
C_v\equiv \frac{\partial E}{\partial T}$ is independent of our choice for $%
U_o$. Thus, although $\xi $ obviously depends on one's choice of $U_o$, $C_v$
does not.

To see what $C_v$ actually measures, let's suppose that we have a quadratic
system whose multiplicity obeys $\Omega =(\frac{U-U_1}{\varepsilon _o})^{%
\frac{\nu N}2}$. It then follows simply that $\frac Sk=\frac{\nu N}2\ln (%
\frac{U-U_1}{\varepsilon _o})$, $\frac{\partial S}{\partial E}=\frac{\nu N}{%
2(U-U_1)}$, $\frac{U-U_1}{kT}=\frac{\nu N}2=\frac{C_v}k$. Thus $\frac{C_v}k$
estimates not $\frac U{kT}$ but $\frac{U-U_1}{kT}$, where $U_1$ is the
``true origin'' of thermal energy for this quadratic system. This is illustrated 
in Figure 2, which plots for an ideal monatomic gas (the classical 
quadratic system) the same quantities plotted for water in Fig. 1. If the system
is not simply quadratic (e.g. if it has phase changes, modes of energy
storage which freeze out, etc.), then $\frac{C_v}k$ is simply a local
estimate of thermal energy over $kT$, under the quadratic assumption. Thus $%
\frac{C_v}k$ modifies the log-log derivative of multiplicity with respect to
energy, by combining it with its rate of increase per e-fold change in
temperature, to yield an estimate of $\frac{U-U_1}{kT}$, where $U_1$ is a
zero of thermal energy determined by assuming that the locally-measured
log-log derivative of multiplicity is constant down to $T=0$.  

This observation also provides a different perspective on the mechanism by
which heat capacity blows up during a phase change. Thermal energy over $kT$
(or the log-log derivative of multiplicity) of course should have no
singularities in it, since both energy and $\frac 1{kT}$ are expected to be
finite for finite systems. It is thus the 2nd term in the two-term
expression for heat capacity above, namely the temperature derivative, that
provides the instability. Discontinuous shifts in the locally-inferred
energy zero ($U_1$) on which heat capacity is based during a phase change
thus, via this second term, also cause heat capacity to become singular.

Returning to Figure 1, this raises the interesting question: {\em Does the
thermal energy of steam increase, decrease, or go negative, when it
condenses to water?} From the above, we can see that it of course decreases
(perhaps even goes negative) if one's ``zero of thermal energy'' is held
constant, since steam loses the latent heat of vaporization when it
condenses, bringing it's total energy down. However, because the specific
heat of water at boiling is higher than that for steam, energy of random
motion measured with respect to our locally-inferred zero of thermal energy
(e.g. at 100C) actually goes up! In other words, a small part of the binding
energy, liberated when water molecules fall into the potential well of their
neighbors, goes to {/em increase} the energy of random motion in the condensed
phase relative to that available to particles in
uncondensed gas!

\section{Beyond Equipartition}

The observations above suggest that introductory texts might consider
highlighting $\frac{C_v}k$ per molecule in natural units for common
substances, with no apology for the fact that it is near but not exactly
half-integral in many cases. After all, thermal energy may not have 
the same access to all molecules, all of the time.  This quantity 
nonetheless provides deep insight into the
relationship between uncertainty and thermal energy.

In shifting the focus from historical temperature units to the
multiplicities which underlie our inferences, we can say that both $\xi $
and $\frac{C_v}k$ measure bits of uncertainty per 2-fold increase in energy
for {\em all} physical systems, with respect to their respective choices for
energy origin. In this sense, they represent physical quantities like
degrees freedom, but with wider applicability. After all, {\em degrees
freedom} presumes not only multiplicities that are linear with energy on a
log-log plot (like a high-temperature Einstein solid), but it also presumes
quadratic energies (i.e. energies proportional to a sum of squares of some
``randomly-occupied'' coordinates of state). The idea that ``every active
coordinate gets $\frac{kT}2$'', as we show below, even more strongly resists
extension to systems with one or more entropy maxima. We begin, however,
with a non-quadratic example of less drastic proportion.

\subsection{Debye Solids}

The Debye heat capacity of a solid is one case where $\frac{C_v}k$ depends
strongly on temperature\cite{Girifalco}. In the Debye low temperature limit,
one has $E=\frac{6N}2\frac{\pi ^4T^3}{5\theta ^3}kT$, so that $\xi =\frac{6N}%
2\frac{\pi ^4T^3}{5\theta ^3}$, while $\frac{C_v}k=\frac{6N}2\left( \frac{%
\pi ^4T^3}{5\theta ^3}+T\frac{3\pi ^4T^2}{5\theta ^3}\right) =4\frac{6N}2%
\frac{\pi ^4T^3}{5\theta ^3}$. Here of course, $\theta $ is the Debye
temperature related to the density and speed of sound in the solid. Note
that in this limit, only a quarter of $\frac{C_v}k$comes from equipartition (%
$\frac E{kT}$), the remaining three quarters from the time derivative of $%
\frac E{kT}$ (in effect, from the unfreezing of new modes of energy
accomodation). As you can see from Figure 1, such unfreezing is associated
with a lowering of the thermal energy zero locally referenced by the heat
capacity. Thus attempts to infer $\xi $ from the heat capacity by assuming
that $\xi \simeq \frac{C_v}k$ yield a 4-fold overestimate of the number of
degrees of freedom! This overestimate decreases as temperatures work
themselves up to and beyond the Debye temperature $\theta $ of the solid, as
illustrated in Figure 1 for temperatures well above $\theta $. There in the
high temperature limit, $\xi \simeq \frac{C_v}k\simeq 3N$, as one expects
from a classical lattice model above with $\Omega \propto E^{3N}$.

\subsection{Two-State Paramagnets}

A system more challenging to the traditional interpretation of ``degrees
freedom'' is that for a system of $N$ half-integral spins (i.e. a two-state
paramagnet) of orientation energy $\varepsilon $. I like it because, as Dan
Schroeder says\cite{Schroeder}, ``it forces us to think primarily in terms
of entropy rather than temperature''. Begin with any system whose
energy-storing coordinates (e.g. displacements in a potential field) in
practice have an upper limit on the amount of energy they'll accomodate. As
long as energy is low enough that no single coordinate approaches the
maximum value, then we may well find behavior very much like that of the
systems discussed above.

However, when individual coordinate energies begin to approach their maximum
(this happens quickly for two-state paramagnets whose coordinates accomodate
but one unit of energy), things change fundamentally. In particular, there
will be but one way (neglecting degeneracies) for the system to store the
maximum amount of energy (namely when each of the coordinates is fully
energized). With but one way to accomodate either minimum energy or maximum
energy, multiplicities will approach 1 at both endpoints of the continuum.
Since large systems may have many ways to store intermediate amounts of
energy, multiplicity (and entropy) as a function of system energy will have
a maximum (or {\em maxima}) somewhere between. At such maxima, $\frac{dS}{dE}%
\equiv \frac 1{kT}$ will be zero, while on their high-energy side $\frac{dS}{%
dE}\equiv \frac 1{kT}$ will be negative (signaling ``population-inverted''
states not accessible by thermal contact with reservoirs at positive
absolute temperature).

Taylor-expanding to second order about the energy ($E_{S\max }$) of such
multiplicity maxima gives $S(E)$ in the neighborhood as $S(E_{S\max })+\frac 
12A\left( E-E_{S\max }\right) ^2$, where $A\equiv \left( \frac{d^2S}{dE^2}%
\right) _{E=E_{S\max }}<0$. Hence for energies near $E_{S\max }$, $\frac{dS}{%
dE}\equiv \frac 1{kT}\simeq A\left( E-E_{S\max }\right) $. Thus deviations
from $E_{S\max }$ are negative at positive absolute temperature, and (for
small deviations at least) are proportional to reciprocal temperature $\beta
\equiv \frac 1{kT}$, as illustrated in Figure 3. Of course, this law was 
discovered by Pierre Curie in experimental study of magnetization, and bears 
his name.

For two-state paramagnets, $A=-\frac 1{N\varepsilon _B^2}$ and $E_{S\max }=%
\frac{N\varepsilon _B}2$, where $\varepsilon _B$ is the energy of alignment
per spin (magnetic moment times magnetic field strength). Thus for $E$ near $%
E_{S\max }$, $\frac{Cv}k\simeq N\varepsilon _B^2\beta ^2>0$, while $\frac{%
E-E_{S\max }}{kT}\simeq -N\varepsilon _B^2\beta <0$. Since $\beta $ for
these systems may be positive or negative, $\xi $ will be negative for some $%
\beta $ values regardless of our choice of the thermal energy zero! Although
negative ``degrees freedom'' may cause discomfort for some, a negative value
for $\xi $ should disturb no one since, to paraphrase a related comment by
Schroeder\cite{Schroeder}, there's no law of physics guaranteeing that there
will not be fewer ways to distribute energy, as more energy is added.

In fact, as we now know, temperature and reciprocal temperature are simply
different forms for the Lagrange multiplier that characterizes a system's
willingness to share thermal energy\cite{Jaynes,Plischke}. Systems, like
these spin systems, capable of taking on (and sharing energy from) negative
absolute temperature states show clearly that for them reciprocal
temperature has more fundamental significance, and that the ``absolute
zeros'' of temperature (approached from negative or positive directions) are
indeed at opposite ends of a continuum\cite{Castle}. But if reciprocal
temperature is more fundamental, our instantaneous no-work heat capacity
should be no less simply connected to the log-log derivative via reciprocal
temperature. Rearrangement of the equation above shows that indeed this is
the case.

\begin{equation}
\frac{C_v}k=-\beta ^2\frac{\partial E}{\partial \beta }=-\beta ^2\frac 
\partial {\partial \beta }\left[ \left\{ \frac{\partial \ln \Omega }{%
\partial \ln E}\right\} \frac 1\beta \right] =\left[ 1-\beta \frac \partial {%
\partial \beta }\right] \left\{ \frac{\partial (\ln \Omega )}{\partial
\left( \ln E\right) }\right\} =-\beta \frac{\partial \ln \Omega }{\partial
\beta }=-\beta \frac{\partial (S/k)}{\partial \beta }\text{.}
\end{equation}
Had we historically adopted as our measure of ``willingness to share
energy'' some other power of the uncertainty slope, say $\gamma \equiv \beta
^a$ where $a\neq -1$, the instantaneous heat capacity would have remained 
proportional to the log-log derivative of multiplicity with respect to 
that measure as well. 

This version of the relation now lets us simplify our perspective on 
entropy maxima.  In the continuum (Stirling) approximation for spin system 
accessible states $%
\Omega $, and measuring energy from the ``low-energy side'', equation (\ref
{temperature}) yields $E\simeq N\varepsilon /(1+e^x)$ where $x\equiv
\varepsilon /kT$, so that $\frac E{kT}=Nx/(1+e^x)$ and $\frac{C_v}k%
=[x/(1+e^x)]^2e^xN$. Here of course, $\varepsilon /kT$ takes on positive and
negative values, ranging from around $+\ln (N)$ to $-\ln (N)$ respectively
for orientation energies $E$ with allowed values from $0$ to $N\varepsilon $%
. As you can see from the plot in Fig. 3, again $\frac{C_v}k$ overestimates $%
\frac E{kT}$ at low temperatures (high values of $\varepsilon /kT$),
although the estimate becomes exact when $\varepsilon /kT$ decreases to
around $x\equiv \frac \varepsilon {kT}\cong 1.279$ (the solution of $%
e^{-x}=x-1$). After this $\frac{C_v}k$ {\em underestimates} $\frac E{kT}$,
which begins to decrease as $x$ decreases and $T$ increases from this point.

All of this switches again when $\frac \varepsilon {kT}$ passes through
zero, since $\frac{C_v}k$ remains positive while the change in uncertainty
per $e$-fold increase in energy ($\frac E{kT}\equiv \xi $) becomes negative
since uncertainty about the system state {\em decreases} with added energy
past this point. Also, of course, average heat capacity goes to zero and no
longer equals $\frac E{kT}$, since the average must be obtained piecewise
when temperature (unlike reciprocal temperature) breaches the discontinuity
from plus infinity to minus infinity. Thus in addition to information units
for heat capacity, we gain from this approach a way to visualize the limits 
of equipartition, and minimize consternation over negative degrees of 
freedom (e.g. for spin and virial systems) as well.

\section{Summary}

In short, we've looked here at natural (as distinct from historical) units
for the common thermodynamic quantities, so that we might explore the
possibility that common uses of $T$ and $1/T$, as measures of ``willingness
to share thermal energy'', have inherited their present emphasis partly
because they predate our present understanding of multiplicity (the $W$ in $%
S=k\ln W$ on Josiah Willard Gibb's tombstone). Heat capacity is a
particularly knotty concept in this regard, since for most of us it has
always been a change in energy ``per degree Kelvin''. In fundamental units,
if heat capacity has any dimensions at all they are those of the base-$b$
information units, since ``change in energy per unit change in energy per
bit'' leaves us with nothing but bits in the bargain.

We point out a simple interpretation for the result, namely that thermal
energy $E$, divided by temperature $T\equiv \frac{\partial E}{\partial S}\equiv
\beta ^{-1}$, is fundamentally $\frac{E}{kT} \equiv E\frac{\partial S}{\partial E}$, i.e. 
the log-log derivative of multiplicity with respect to energy (e.g. a
measure of the bits of uncertainty increase per two-fold increase in
energy). The instantaneous no-work heat capacity in this context is $%
\frac{C_v}{k}\equiv T\frac{\partial S}{\partial T}=-\beta \frac{\partial S}{\partial
\beta }$, an estimate of $\frac{E}{kT} $ with an energy-zero inferred from the
``local slope'' of the log-log plot. We show that these two quantities bear
a simple relationship to each other, regardless of the variable (e.g. $T$ or 
$\frac 1{kT}$) chosen to keep track of a system's willingness to share
energy thermally. 

The former of the quantities, namely $\frac{E}{kT} $, plays the
role of ``degrees freedom over two'' in quadratic systems, but is dependent 
on the energy zero, and regardless can take on negative values in
systems with entropy maxima. Its limitations are those of the concept 
of equipartition itself.  The latter quantity, namely $\frac{C_v}k$,
provides deep insight into the ways a system accomodates {\em new} thermal
energy. Because these quantities are defined in terms of state multiplicity
and the conserved variable being shared (in this case energy), and relatively 
independent of the form chosen for the Lagrange multiplier in the problem
(e.g. temperature in historical units), their {\em analogs} in problems that
involve the sharing of other conserved quantities (e.g. volume, particles,
or even dollars) may be easier to recognize and put to use as well.

\acknowledgments

This work has benefited indirectly from support by the U.S. Department of
Energy, the Missouri Research Board, as well as Monsanto and MEMC Electronic
Materials Companies. It has benefited most, however, from the interest and
support of students at UM-St. Louis.

\begin{figure}[tbp]
\caption{Plot of $E$ versus $kT$ (lower left), $C_v/k$ versus $E/kT$ (upper right), 
and the entropy plots whose slopes associate therewith (see text), for a steam, 
water, and Debye-model ice system ($T_{Debye}=333$K assuming sound speed 
near $3500$[m/s]) at 1 atmosphere.  This plot illustrates the use of physical (as distinct 
from historical) temperature units, and of dimensionless units (base-b information 
units per b-fold increase in T or E) for specific heat $C_v/k$ and multiplicity's 
energy-exponent $E/kT$.  All quantities are ``per molecule'' values.  Note in particular 
how the quadratic model "energy of thermal motion'' for water {\em increases} as water 
condenses from a gas to a liquid.  Also note how equipartition applies only far from 
phase changes and ``freeze out'' zones, and then only if a fictituous zero for the 
thermal energy is chosen.
}
\label{Fig1}
\end{figure}

\begin{figure}[tbp]
\caption{Plot of $E$ versus $kT$ (lower left), $C_v/k$ versus $E/kT$ (upper right), 
and the entropy plots whose slopes associate therewith (see text), for an ideal 
(Sakur-Tetrode) monatomic 6000-atom argon gas at 1 atmosphere.  This figure 
illustrates the form taken by the plots introduced in Fig. 1, for the case of an ideal 
``quadratic system''.  All quantities are ``per atom'' values.  This system 
behaves like many physical systems 
(including low-density gases and Dulong-Petit metals), but only when well away 
from temperatures where excitation modes are freezing out, or 
phase changes are imminent.  On the ``equipartition line'' note the 
single point for $C_v/k = E/kT = 3/2$.  If the zero of energy is 
shifted for this system (e.g. by a phase change or by quantum mechanical freezing out 
of excitation modes), the point stretches into a horizontal line moving 
between $E/kT = 0$ and $E/kT = 3/2$.  This line approaches its limiting value 
(here $E/kT = 3/2$) by moving rightward (as for liquid water in Fig. 1), or 
by moving leftward (as for steam in Fig. 1), if the zero of energy is moved, 
respectively, up or down from its quadratic value.}
\label{Fig2}
\end{figure}

\begin{figure}[tbp]
\caption{Plot of $E$ versus $kT$ (lower left), $C_v/k$ versus $E/kT$ (upper right), 
and the entropy plots whose slopes associate therewith (see text), for a two-state 
paramagnet consisting of 10 non-interacting spins.  This differs from the plots in 
Fig. 1 and 2 in that the two left side plots use $1/kT$ in place of $kT$ because of 
its more natural mapping of inverted and uninverted population states.  
All quantities are ``per spin'' values, and of course the energy units $e_o$ 
depend on both the spin magnetic moment and the magnetic field.  Note in particular 
how the ``equipartition line'' seems even less relevant here, even though heat 
capacity remains positive as reciprocal temperature dips below zero.  For virial 
systems (like a gravity bound gas), negative heat capacities make the concepts of 
equipartition and degrees-freedom even less apropo, even though the relationships 
shown here remain intact.}
\label{Fig3}
\end{figure}

\end{document}